\documentclass[a4paper,11pt]{article}

\usepackage{jheppub}
\usepackage[utf8]{inputenc}
\usepackage{graphicx,cancel,float}
\usepackage{amssymb}
\usepackage{textcomp}
\usepackage{amsmath}
\usepackage{tabularx}
\usepackage{bm}
\usepackage{times}
\usepackage{color}
\usepackage{slashed}
\usepackage{multirow}
\usepackage{verbatim}
\usepackage{epsfig,epstopdf}
\usepackage{subfigure}
\allowdisplaybreaks[4]

\def\lsim{\mathrel{\raise.3ex\hbox{$<$\kern-.75em\lower1ex\hbox{$\sim$}}}}
\def\gsim{\mathrel{\raise.3ex\hbox{$>$\kern-.75em\lower1ex\hbox{$\sim$}}}}

\newcommand{\bew}{\begin{widetext}}
\newcommand{\enw}{\end{widetext}}
\newcommand{\bee}{\begin{equation}}
\newcommand{\ene}{\end{equation}}
\newcommand{\bea}{\begin{eqnarray}}
\newcommand{\ena}{\end{eqnarray}}
\newcommand{\bes}{\begin{subequations}}
\newcommand{\ens}{\end{subequations}}

\def\calm{\mathcal{M}}

\definecolor{orange}{rgb}{1,0.5,0}

\subheader{\it \today}

\title{\Large{\bf Production of Light Dark Particles from Nonlinear Compton Scattering Between Intense Laser and Muon or Proton Beam}}

\author[a]{Tong Li}
\emailAdd{litong@nankai.edu.cn}
\author[b]{Kai Ma}
\emailAdd{kai@xauat.edu.cn}
\author[a]{Man Yuan}
\emailAdd{yuanman@mail.nankai.edu.cn}

\affiliation[a]{School of Physics, Nankai University, 94 Weijin Road, Tianjin 300071, China}
\affiliation[b]{Faculty of Science, Xi'an University of Architecture and Technology, Xi'an, 710055, China}

\abstract{
The laser of an intense electromagnetic field promotes the studies of strong-field particle physics in high-intensity frontier. Particle accelerator facilities in the world produce high-quality muon and proton beams. In this work, we propose the nonlinear Compton scattering to light dark particles through the collision between intense laser pulse and muon or proton beam.
We take light dark photon and axion-like particle as illustrative dark particles. The cross sections of relevant nonlinear Compton scattering to dark photon or axion-like particle are calculated. We also analyze the background processes with missing neutrinos.
The prospective sensitivity shows that the laser-induced process provides a complementary and competitive search of new invisible particles lighter than about 1 MeV.
}

\begin{document}

\maketitle
\setcounter{page}{2}

\newpage

\section{Introduction}
\label{sec:Intro}

In 1951, J.~Schwinger showed that at external field strengths of $E\simeq 1.32\times 10^{18}~{\rm V/m}$, the virtual electron-positron pair
fluctuations in quantum electrodynamics (QED) vacuum are converted
into real electron-positron pairs~\cite{Schwinger:1951nm}. This critical field strength corresponds to an extreme intensity limit of laser, that is $I=2.3\times 10^{29}~{\rm W/cm^2}$. The laser of an intense electromagnetic field strength has a lot of applications in such as atomic physics, nuclear physics and particle physics (see a recent review~\cite{Fedotov:2022ely} and references therein).
One is able to study processes that normally do not occur in vacuum but can be induced under strong fields.
Therefore, intense laser pulses allow for the investigation of rich phenomena at the high-intensity frontier of physics.

The E144 experiment performed at SLAC observed two strong-field processes in the interaction of
an ultra-relativistic electron beam with a terawatt laser pulse in 1990s, i.e., the nonlinear Compton scattering and the nonlinear Breit-Wheeler pair production in high intense laser field~\cite{Burke:1997ew,Bamber:1999zt}. The observation of these two processes promotes the studies of non-perturbative QED and nonlinear effects in QED (see a recent review in Ref.~\cite{Hartin:2018egj}). The nonlinear effects from the higher-order terms in the electron beam scattering become significant when
\begin{eqnarray}
\eta_e\equiv {e\mathcal{E}_0\over \omega m_e} \gtrsim 1\;,
\end{eqnarray}
where $e$ is the electric charge unit, $\mathcal{E}_0$ is the electromagnetic field strength, $m_e$ is the electron mass and $\omega$ is the laser beam energy. Interestingly, these laser-induced scattering processes can provide a potentially useful strategy for the search of new physics beyond the Standard Model (SM)~\cite{Fuchs:2024edo,Dillon:2018ypt,King:2018qbq,Bai:2021gbm,Dillon:2018ouq,King:2019cpj,Beyer:2021mzq,Huang:2020lxo,Ma:2024ywm,Ma:2025axq,Ma:2025ymt}.

Nevertheless, with the help of higher power lasers, the collision of laser pulse and muon or proton beam may lead to interesting phenomena.
For instance, the vacuum polarization correction to elastic muon scattering was explored in an intense field~\cite{Ritus:1972nf}. Muon decay can be significantly altered in a strong electromagnetic field~\cite{1983A,Ritus:1985vta,Narozhny:2008zz,Dicus:2008nw,Farzinnia:2009gg,King:2025ywl}. The laser-muon interaction was also proposed to probe exotic physics beyond the SM~\cite{Tizchang:2016hml,Tizchang:2018mzr}.

Nowadays, high-quality muon and proton beams are being developed to enable studies in high-precision muon physics and high-energy elementary particle physics. Particle accelerator facilities in the world provide the beam sources, for example, Paul Scherrer Institute (PSI) in Switzerland, Fermi National Accelerator Laboratory (FNAL) and Brookhaven National Laboratory (BNL) in the United States, Japan Proton Accelerator Research Complex (J-PARC), China Spallation Neutron Source (CSNS) and so on. For these operations, protons are accelerated
in the linac and booster synchrotron and directed to a target. The secondary beam pions then decay into final muons. For example, in FNAL accelerator complex, protons are accelerated to 8 GeV in booster and $\mu^+$ beam is selected in momentum as 3.1 GeV so as to measure muon anomalous magnetic moment~\cite{PhysRevAccelBeams.22.011001,Muong-2:2023cdq}. Moreover, the proton and muon beams are designed to be accelerated to TeV energy scale for the direct search of new physics beyond the SM at hadron colliders (e.g., Tevatron at FNAL and LHC at CERN) or the proposed muon colliders~\cite{AlAli:2021let,InternationalMuonCollider:2025sys,Hamada:2022mua}. It is intriguing to consider the access of new physics at low-energies by combining increasingly intense lasers and energetic muon or proton beams.

In this work, we investigate the collision of intense laser pulse and a high-energy muon ($\mu^+$) or proton ($\mathcal{P}$) beam. The laser-induced nonlinear Compton scattering can produce a new massive dark particle beyond the SM~\cite{Dillon:2018ypt,King:2018qbq,Dillon:2018ouq,Ma:2024ywm}. We consider two illustrative invisible particles, i.e., dark photon (DP) and axion-like particle (ALP). In the nonlinear Compton scattering, a muon or proton beam absorbs multiple photons from the laser field and radiates a single dark particle
\begin{eqnarray}
\mu^+/\mathcal{P} \to \mu^+/\mathcal{P} + {\rm DP~or~ALP}\;.
\end{eqnarray}
DP couples to SM fermions through the kinetic mixing with visible photon. The fermionic interactions with the derivatively
coupled ALP are at dimension 5. They are expected to be long-lived and escape the detector if their masses are less than $2m_e\approx 1$ MeV. We calculate the nonlinear Compton scattering cross sections of single fermion together with invisible DP and ALP. The charged and neutral weak currents in the SM can also result in single fermion with missing neutrinos in final states. We analyze the relevant SM background processes mediated by $W^\pm$ or $Z$ bosons in details.
Compared to other collider or beam dump experiments, the sizable cross section of this process for the light dark particle creation results in an increased observation sensitivity to the kinetic mixing of DP and ALP derivative coupling to muon or proton.

Note that the major application of laser-proton interaction is laser-driven proton acceleration. Laser-proton colliders are still in the proof-of-principle stage for facilities such as Extreme Light Infrastructure--Nuclear Physics (ELI--NP)~\cite{ELINP}. Their main challenges include insufficient proton beam quality, low collision luminosity, and low signal-to-noise ratios in diagnostic systems. The demonstration in laser-proton collision physics will require more advanced beam control technology and laser systems with higher repetition rates.

This paper is organized as follows. In Sec.~\ref{sec:LaserCompton}, we review the dark particle hypotheses of DP and ALP. The basics of laser-induced nonlinear Compton scattering are also discussed in details. In Sec.~\ref{sec:DPALP}, we consider the laser-induced Compton scattering to dark particles. The cross sections of the relevant Compton scattering to DP or ALP are calculated. We also analyze the SM background processes with missing neutrinos. We show the sensitivity reach for dark particle parameters in Sec.~\ref{sec:results}. Our conclusions are drawn in Sec.~\ref{sec:Con}.

\section{Dark particle hypotheses and laser-induced scattering}
\label{sec:LaserCompton}

\subsection{Dark particle hypotheses: dark photon and axion-like particle}

The dark photon (or called hidden photon)~\cite{Holdom:1985ag,Holdom:1986eq} is a spin-one particle gauged by an Abelian group $U(1)_{\rm D}$ in dark sector (see a recent review Ref.~\cite{Fabbrichesi:2020wbt} and references therein). A gauge kinetic
mixing between the field strength tensors of the SM electromagnetic gauge $U(1)_{\rm EM}$ group and the $U(1)_{\rm D}$ group is introduced.
The relevant DP Lagrangian is
\begin{eqnarray}
\mathcal{L}_{\rm DP}\supset -\frac{1}{4}F^{\mu\nu}F_{\mu\nu}-\frac{1}{4}{F}_D^{\mu\nu}{F}_{D\mu\nu}-\frac{\epsilon}{2}F^{\mu\nu}{F}_{D\mu\nu}+\frac{1}{2}m_{D}^2{A_D}^\mu A_{D\mu}\;,
\end{eqnarray}
where $F^{\mu\nu}$ ($F^{\mu\nu}_D$) is the SM (dark) electromagnetic field strength, and $A_D$ is the dark gauge boson with mass $m_{\gamma_D}$. By assuming that the SM particles are uncharged under the dark gauge group, the kinetic mixing $\epsilon$ can be obtained by integrating out heavy particles charged under both gauge groups at one-loop level. The two gauge fields can then be rotated to get rid of the mixing. As a result, the SM matter current gains a shift as $A_\mu\to A_\mu - \epsilon A_{D\mu}$. Then, the physical DP $\gamma_D$ has a coupling to the electromagnetic current of the SM fermions
\begin{eqnarray}
-e \epsilon J_{\rm EM}^\mu A_{D\mu}=-eQ\epsilon \overline{\psi}\gamma^\mu \psi A_{D\mu}\;.
\end{eqnarray}
The massive DP particle can be searched in terrestrial experiments in the presence of this kinetic mixing.

The axion-like particle is a CP-odd pseudo-Nambu-Goldstone boson from the spontaneous breaking of a global $U(1)$ symmetry. The QCD axion is a well-considered example of such CP-odd particle~\cite{Peccei:1977hh,Peccei:1977ur,Weinberg:1977ma,Wilczek:1977pj} (see a recent review Ref.~\cite{DiLuzio:2020wdo}).
The mass of generic ALP ($m_a$) and the symmetry breaking scale (or called decay constant $f_a$) associated with an ALP can be completely free parameters in many other theoretical models~\cite{Dimopoulos:1979pp,Tye:1981zy,Zhitnitsky:1980tq,Dine:1981rt,Holdom:1982ex,Kaplan:1985dv,Srednicki:1985xd,Flynn:1987rs,Kamionkowski:1992mf,Berezhiani:1999qh,Berezhiani:2000gh,Hsu:2004mf,Hook:2014cda,Alonso-Alvarez:2018irt,Hook:2019qoh}. The range of ALP mass spans from sub-micro-eV~\cite{Kim:1979if,Shifman:1979if,Dine:1981rt,Zhitnitsky:1980tq,Turner:1989vc} to TeV scale and even beyond~\cite{Rubakov:1997vp,Fukuda:2015ana,Gherghetta:2016fhp,Dimopoulos:2016lvn,Chiang:2016eav,Gaillard:2018xgk,Gherghetta:2020ofz}. Thus, the search for the ALPs relies on rather different strategies and facilities in experiments. The probe of ALP also depends on the interactions between ALP and the SM fields. Here, we are interested in the ALP-fermion coupling defined via the Lagrangian term
\begin{eqnarray}
\mathcal{L}_{\rm ALP}\supset c_{af} {\partial_\mu a\over 2f_a} \overline{f} \gamma^\mu \gamma_5 f\;,
\end{eqnarray}
where $c_{af}$ is a dimensionless constant and $f=\mu, \mathcal{P}$ here. One usually defines another dimensionless ALP-fermion coupling $g_{af}=c_{af}m_f/f_a$.
In this work, we consider proton-philic ALP or muon-philic ALP scenario. As a phenomenological analysis, we remain agnostic about the origin of the fermion-philic ALP. The ALP-fermion couplings are simply taken as independent parameters in an effective framework. Flavor non-universality is expected in the most general effective Lagrangian for fermionic ALP interactions~\cite{Georgi:1986df,Bauer:2021mvw} because non-universal Peccei-Quinn (PQ) charge matrices can be embedded to the mass eigenstate basis. An example is given by models relating the PQ symmetry $U(1)_{\rm PQ}$ to a
global $U(1)$ flavor symmetry~\cite{Ema:2016ops,Calibbi:2016hwq,Bjorkeroth:2018ipq,delaVega:2021ugs}. The muon-philic ALP would emerge by assigning specific PQ charge for leptons. Suppose ALP-photon coupling is induced, given the strong bound on ALP-photon coupling and the ALP mass region of interest in this work $m_a< 1~{\rm MeV}\approx 2m_e$, the emitted light ALP would be long-lived enough as missing energy.

\subsection{Basics of laser-induced nonlinear scattering}

In the presence of an electromagnetic potential, the wave function of a relativistic fermion of mass $m$ subjects to the following Dirac equation
\bee
(i \slashed{\partial} - Qe \slashed{A} -m) \psi(x)=0 \;,
\ene
where $e$ is the unit of electric charge and $Q$ is the charge operator (e.g. $Q\psi=+\psi$ for proton).
With circular polarization, the vector potential $A^\mu$ takes the following form
\bee
\label{eq:laser}
A^\mu(\phi)= a_1^\mu \cos (\phi)+ a_2^\mu \sin (\phi) \,,\quad \phi=k \cdot x,
\ene
where $k$ is the four-momentum of laser photon with $k^0=\omega$ and
$\phi$ is the phase of the laser field. The polarization vectors are given by
\bea
a_1^\mu = |\vec{a}\,|(0,1,0,0)\,,~~~
a_2^\mu = |\vec{a}\,|(0,0,1,0)\;.
\ena
They satisfy the following relations
\bea
a_1 \cdot a_2 = 0\;,~~~
a_1^2 = a_2^2 = -|\vec{a}\,|^2 \equiv -a^2\;,
\ena
where $a=\mathcal{E}_0/\omega$ with $\mathcal{E}_0$ being the electric and magnetic field strength.
Note that here we choose the laser field to be circularly polarized and monochromatic~\cite{King:2018qbq}.
In fact, the spatial dependence of the vector potential is described by a function $f(x)$ for the pulse shape in a realistic experiment~\cite{Dillon:2018ypt}. The pulse shape for $f(x)$ is usually assumed to be Gaussian with respect to the phase $\phi$. The spatial dependence of the vector potential would enter the following calculation of the $S$ matrix element through the Fourier transform of the pulse shape $\tilde{f}$.
Next we instead work in the monochromatic limit.

The wave functions of the incoming proton $\mathcal{P}$ and anti-muon $\mu^+$ with $Q=+1$ charge are given by the Volkov state~\cite{Wolkow1935}
normalized to the volume $V$ as follows
\begin{eqnarray}
\psi_{p, s}(x)
&=&
\left[1+\frac{Qe \slashed{k} \slashed{A}}{2\left(k \cdot p\right)}\right]
\frac{u\left(p, s\right)}{\sqrt{2 q^0 V}} e^{i F(q, s)}\;,\\
\overline{\psi}^{(+)}_{p, s}(x)
&=&
\frac{\overline{v}\left(p, s\right)}{\sqrt{2 q^0 V}}\left[1-\frac{Qe \slashed{k} \slashed{A}}{2\left(k \cdot p\right)}\right]
 e^{i F(q, s)}\;,
\end{eqnarray}
where the label ``$(+)$'' refers to the wave function of anti-particle, $u(p, s)$ and $v(p, s)$ are the usual Dirac spinors for the free particle and anti-particle, respectively, and $p$ denotes the initial momentum before entering the electromagnetic background. In the external plane-wave field, they receive an ``effective momentum''
\bee
q^\mu
=
p^\mu + \frac{Q^2 e^2 a^2}{2 k \cdot p} k^\mu
\ene
which satisfies the following dispersion relation
\bee
q^2 = m^2_f + Q^2 e^2 a^2 = m_f^{\ast 2}\;,~~~f=\mu, \mathcal{P} \,.
\ene
The phase function $F_1(q, s)$ is given by
\bee
F(q, s)
=
-q \cdot x - \frac{Qe\left(a_1 \cdot p\right)}{\left(k \cdot p\right)} \sin \phi
+
\frac{Qe\left(a_2 \cdot p\right)}{\left(k \cdot p\right)} \cos \phi \;.
\ene
Similarly, the Volkov state of the final proton or $\mu^+$ in the laser field can be obtained with the substitution $p\to p'$ and $q\to q'$. The Volkov wave function is the exact solution of the Dirac equation describing the ``dressed'' fermion which continuously
interacts with the traveling cloud of laser photons. It automatically includes the nonlinear effects of the laser field in the series of $ea/m_f$.

\section{Laser-induced nonlinear Compton scattering to dark particles}
\label{sec:DPALP}

\subsection{Nonlinear Compton scattering to dark particles}

In this section, we start from the calculation of laser-induced Compton scattering to new massive particles in intense laser field. As illustrative examples, we consider dark photon $\gamma_D$ and ALP $a$ as dark particles in the following Compton scattering processes
\begin{eqnarray}
\mu^+/\mathcal{P}(p) + n\omega (k) \to \mu^+/\mathcal{P}(p') + \gamma_D/a (k')\;,
\end{eqnarray}
where $n$ denotes the number of optical photons with energy $\omega$.

The lowest-order scattering $S$ matrix element for the laser-proton production of DP $\gamma_D$ reads
\begin{eqnarray}
S_{f i}^{\mathcal{P}\gamma_D}
&=&
ie\epsilon {1\over \sqrt{2k^{\prime 0} V}}\int d^4 x e^{ik'\cdot x}\; \overline{\psi}_{p', s'}(x) \cancel{\varepsilon}_D \psi_{p, s}(x) \;,
\end{eqnarray}
where $\varepsilon^\mu_D$ is the polarization vector of the outgoing physical DP.
The scattering amplitude is
\begin{eqnarray}
\label{eq:am}
\calm^{\mathcal{P}\gamma_D}
&=&ie \epsilon {1\over \sqrt{2k^{\prime 0} V}}e^{ik'\cdot x}\overline{\psi}_{p', s'}(x) \cancel{\varepsilon}_D \psi_{p, s}(x)\nonumber\\
&=&
ie \epsilon {e^{i(k'+q'-q)\cdot x}e^{-i\Phi}\over \sqrt{2^3 V^3 q^0 q^{\prime 0} k^{\prime 0}}} \overline{u}(p', s') \left[1+\frac{e \slashed{A}\slashed{k}  }{2 k \cdot p'}\right]\cancel{\varepsilon}_D \left[1+\frac{e \slashed{k} \slashed{A} }{2 k \cdot p}\right] u(p,s)\;,
\end{eqnarray}
where the phase factor $\Phi$ is given by
\bee
\Phi =
-e a_1 \cdot y \sin \phi + e a_2 \cdot y \cos \phi
\ene
with $y$ being defined as
\bee
y^\mu = \frac{ p^{\prime \mu} }{k \cdot p'} - \frac{ p^\mu }{k \cdot p}\;.
\ene
After substituting the electromagnetic field given in Eq.~\eqref{eq:laser}, the amplitude can be factorized as follows
\begin{eqnarray}
\calm^{\mathcal{P}\gamma_D}
&=&
ie \epsilon {e^{i(k'+q'-q)\cdot x}\over \sqrt{2^3 V^3 q^0 q^{\prime 0} k^{\prime 0}}}
\sum_{i=0}^{2}
C_i \calm_i^{\mathcal{P}\gamma_D}\;.
\end{eqnarray}
The reduced amplitudes $\calm_i^{\mathcal{P}\gamma_D}$ become
\bea
\calm_0^{\mathcal{P}\gamma_D}
&=&
\overline{u_2} \cancel{\varepsilon}_D u_1 +\overline{u_2}{e^2 a^2\over 2k\cdot p k\cdot p'} k\cdot \varepsilon_D \cancel{k}u_1\;,
\\[3mm]
\calm_1^{\mathcal{P}\gamma_D}
&=&
+\overline{u_2}\cancel{\varepsilon}_D
\frac{e \slashed{k} \slashed{a}_1 }{2 k \cdot p} u_1 + \overline{u_2}
\frac{e \slashed{a}_1 \slashed{k} }{2 k \cdot p'} \cancel{\varepsilon}_D u_1\;,
\\[3mm]
\calm_2^{\mathcal{P}\gamma_D}
&=&
+\overline{u_2}\cancel{\varepsilon}_D
\frac{e \slashed{k} \slashed{a}_2 }{2 k \cdot p} u_1 + \overline{u_2}
\frac{e \slashed{a}_2 \slashed{k} }{2 k \cdot p'} \cancel{\varepsilon}_D u_1\;,
\ena
where $u_1\equiv u(p,s)$ and $\overline{u_2}\equiv \overline{u}(p',s')$.
The corresponding coefficients $C_i$ are given by
\bea
C_0 &=& e^{-i\Phi}\;,
\\[3mm]
C_1 &=& \cos\phi \,e^{-i\Phi}\;,
\\[3mm]
C_2 &=& \sin\phi \,e^{-i\Phi}\;,
\ena
where the phase $\Phi$ takes the form
\bee
\Phi  =  z \sin(\phi - \phi_0)
\ene
with
\bee
z = e \sqrt{ (a_1 \cdot y)^2 + (a_2\cdot y)^2 } \,,
\quad
\cos\phi_0 =  -\frac{e a_1 \cdot y }{ z} \,,
\quad
\sin\phi_0 =  -\frac{e a_2 \cdot y}{ z} \,.
\ene

Since the function $e^{-i\Phi}$ is a periodic function, it can be performed by Fourier series expansion as
\bee
e^{-i\Phi}
=
e^{- i z \sin \left(\phi-\phi_0\right)}
=
\sum_{n=-\infty}^{\infty} c_n e^{-\mathrm{i} n\left(\phi-\phi_0\right)}\;,
\ene
where the expansion coefficients are given as
\bee
c_n
=
\frac{1}{2 \pi} \int_{-\pi}^\pi \mathrm{d} \varphi \, e^{-\mathrm{i} z \sin\varphi} e^{\mathrm{i} n \varphi}
=J_n(z)\;.
\ene
This is the integral representation of the Bessel function $J_n(z)$. Thus, we have
\bee
C_0=e^{-i\Phi} = \sum_{n=-\infty}^{\infty} B_n(z) e^{-\mathrm{i} n \phi}\;,
\ene
where $B_n(z)=J_n(z) \mathrm{e}^{\textcolor{cyan}{-}i n \phi_0}$. This is a well-known relation known as Jacobi-Anger expansion.
By using the following relations
\bea
\cos\phi = \frac{1}{2} \Big( e^{i\phi} + e^{-i\phi} \Big) \,,~~~
\sin\phi = \frac{1}{2i} \Big( e^{i\phi} - e^{-i\phi} \Big) \,,
\ena
we obtain
\bea
C_1=
\cos\phi \,e^{-i\Phi}
&=&
\frac{1}{2}\sum_{n=-\infty}^{\infty} B_n(z)  \Big[ e^{-i(n-1)\phi} + e^{-i(n+1)\phi} \Big]\;,
\\[3mm]
C_2=
\sin\phi \,e^{-i\varPhi}
&=&
\frac{1}{2i}\sum_{n=-\infty}^{\infty} B_n(z)  \Big[ e^{-i(n-1)\phi} - e^{-i(n+1)\phi} \Big]\;.
\ena
Given the redefinition of integers $n' = n+1$ and $n'' = n-1$, together with the completeness of the $n'$ and $n''$, they become
\bea
C_1=
\cos\phi \,e^{-i\varPhi}
&=&
\frac{1}{2}\sum_{n=-\infty}^{\infty} \Big[ B_{n+1}(z) + B_{n-1}(z)  \Big] e^{-in\phi} \;,
\\[3mm]
C_2=
\sin\phi \,e^{-i\varPhi}
&=&
\frac{1}{2i}\sum_{n=-\infty}^{\infty}  \Big[ B_{n+1}(z) - B_{n\-1}(z)  \Big] e^{-in\phi} \;.
\ena

Then, the amplitude can be simplified as
\bee
\calm^{\mathcal{P}\gamma_D}
=
ie \epsilon \sum_{n=-\infty}^{\infty}
\frac{ e^{i( k' +q' - q - nk) \cdot x }  }{ \sqrt{2^3 V^3 q^0 q^{\prime 0} k^{\prime 0} }}
\sum_{i=0}^{2}
\widetilde{C}_i^{n} \calm_i^{\mathcal{P}\gamma_D}\;,
\ene
where
\bee
\widetilde{C}_i^{n} = e^{in\phi} C_i^{n} \;,~~~\sum_{n=-\infty}^{\infty}\widetilde{C}_i^{n} e^{-in\phi}=\sum_{n=-\infty}^{\infty}C_i^n=C_i\;.
\ene
After integrating out the coordinate $x$, we have
\bee
S_{f i}^{\mathcal{P}\gamma_D}
=
ie \epsilon \sum_{n=-\infty}^{\infty}
\frac{ (2\pi)^4 \delta^4( k' + q' - q - nk) }{ \sqrt{2^3 V^3 q^0 q^{\prime 0} k^{\prime 0} }}
\sum_{i=0}^{2}
\widetilde{C}_i^{n} \calm_i^{\mathcal{P}\gamma_D}\;,
\ene
and its square is given by
\bee
\big| S_{f i}^{\mathcal{P}\gamma_D}\big|^2
=
e^2 \epsilon^2 \sum_{n=-\infty}^{\infty}
\frac{ (2\pi)^4 \delta^4( k' + q' - q - nk) VT }{ 2^3 V^3 q^0 q^{\prime 0} k^{\prime 0} }
\sum_{i,j=0}^{2}
\widetilde{C}_i^{n} \big( \widetilde{C}_j^{n} \big)^\dag  \overline{ \calm_i^{\mathcal{P}\gamma_D} \calm_j^{\mathcal{P}\gamma_D\dag} }\;.
\ene
The ``overline'' in above equation denotes the summation and average over the spin degree of freedom of relevant particles. For the Compton scattering, after averaging the proton spin, the squared amplitude is then
\begin{eqnarray}
\sum_{i,j=0}^{2}
\widetilde{C}_i^{n} \big( \widetilde{C}_j^{n} \big)^\ast  \overline{     \calm^{\mathcal{P}\gamma_D}_i \calm^{\mathcal{P}\gamma_D\dag }_j}&=& -2(m_\chi^2+2m_{\mathcal{P}}^2)J_n^2(z) \nonumber \\
&&+{e^2a^2(1+u^2)\over u} [ J_{n-1}^2(z)+J_{n+1}^2(z)-2J_{n}^2(z) ] \;,
\label{eq:amp2DP}
\end{eqnarray}
where $m_\chi$ denotes the universal mass of dark particle here and below. Here, we have
\begin{eqnarray}
u & \equiv & {k\cdot p\over k\cdot p'} = \frac{\sqrt{s_{n}}}{Q^{\prime} - |\vec{q}'|
\cos\theta} \;, \nonumber \\ [0.5em]
s_{n} & = & m_{\mathcal{P}}^{*2} + 2n\omega(E_{\mathrm{Lab}} + \cos\theta_{\rm Lab}\ p_{\mathrm{Lab}})\;, \nonumber \\ [0.5em]
Q^{\prime} & = & \frac{s_{n} + m_{\mathcal{P}}^{*2} - m_{\chi}^{2}}{2\sqrt{s_{n}}}\;,~~ |\vec{q}'| = {\lambda^{1/2}(s_n,m_{\mathcal{P}}^{*2},m_{\chi}^{2})\over 2\sqrt{s_n}}\;,
\label{eqn-sn}
\end{eqnarray}
where $\lambda(x, y, z) = x^2+y^2+z^2-2xy-2xz-2yz$, $\sqrt{s_n}$ is the center-of-mass (c.m.) energy, $\theta$ is the scattering polar angle in the c.m. frame, $E_{\rm Lab}$ is the energy of proton beam and $\theta_{\rm Lab}$ is the initial scattering angle in the laboratory frame.

The amplitude of Compton scattering to ALP can be expressed in terms of its ALP-proton coupling
\begin{eqnarray}
\mathcal{M}^{\mathcal{P}a} &=& \frac{c_{a \mathcal{P}}}{2f_{a}} \frac{1}{\sqrt{2{k^{\prime 0}} V}} e^{i{k^{\prime}} \cdot x} \overline{\psi}_{p^{\prime}, s^{\prime}}(x) \cancel{k}^{\prime} \gamma_{5} \psi_{p,s}(x) \nonumber \\
&=& \frac{c_{a \mathcal{P}}}{2f_{a}} \frac{e^{i(k^{\prime}+q^{\prime}-q) \cdot x}e^{-i\Phi}}{\sqrt{2^{3} V^{3} q^{0} q^{\prime 0} k^{\prime 0}}} \overline{u}(p^{\prime}, s^{\prime}) \left[ 1 + \frac{eA\cancel{k}}{2k\cdot p^{\prime}} \right] \cancel{k}^{\prime} \gamma_{5} \left[ 1 + \frac{e \cancel{k} A}{2k \cdot p} \right] u(p,s)\;.
\end{eqnarray}
The spin averaged and squared matrix element for this process is
\begin{eqnarray}
\sum_{i,j=0}^{2}
\widetilde{C}_i^{n} \big( \widetilde{C}_j^{n} \big)^\ast  \overline{ \calm_i^{\mathcal{P}a} \calm_j^{\mathcal{P}a\dag} }&=& -4m_{\mathcal{P}}^2m_\chi^2J_n^2(z) \nonumber \\
&&+{e^2a^2 2 m_{\mathcal{P}}^2(1-u)^2\over u} [ J_{n-1}^2(z)+J_{n+1}^2(z)-2J_{n}^2(z) ] \;.
\label{eq:amp2ALP}
\end{eqnarray}
For laser-muon scattering to DP and ALP, the $S$ matrix elements are respectively given by
\begin{eqnarray}
S_{f i}^{\mu^+\gamma_D}
&=&
-ie\epsilon {1\over \sqrt{2k^{\prime 0} V}}\int d^4 x e^{ik'\cdot x}\; \overline{\psi}_{p, s}^{(+)}(x) \cancel{\varepsilon}_D \psi_{p', s'}^{(+)}(x) \;,
\end{eqnarray}
and
\begin{eqnarray}
S_{f i}^{\mu^+a}
&=&
\frac{c_{a \mu}}{2f_{a}} {1\over \sqrt{2k^{\prime 0} V}}\int d^4 x e^{ik'\cdot x}\; \overline{\psi}_{p, s}^{(+)}(x) \cancel{k}' \gamma_5 \psi_{p', s'}^{(+)}(x) \;.
\end{eqnarray}
We find that their squared amplitudes are the same as their respective results in laser-proton scattering but with the mass substitution $m_\mathcal{P}\to m_\mu$.
The cross section of Compton scattering is given by~\cite{Greiner:1992bv}
\begin{eqnarray}
\sigma &=& {|S_{fi}|^2\over VT}{1\over 2(1/V)}{1\over \rho_\omega} V \int \frac{d^3 q'}{(2 \pi)^3} V \int \frac{d^3 k'}{(2 \pi)^3}\nonumber\\
&=&{1\over 2\rho_\omega}\frac{ \kappa^2 }{2 q^0 } \sum_{n=-\infty}^{\infty}  \int d\Pi_2
\sum_{i,j=0}^{2}
\widetilde{C}_i^{n} \big( \widetilde{C}_j^{n} \big)^\dag
\overline{\calm_i \calm_j^\dag} \;,
\label{eq:xsec}
\end{eqnarray}
where $\kappa=e\epsilon$ ($c_{a\mu,a\mathcal{P}}/(2f_a)$) for DP (ALP), $S_{fi}$ ($\calm$) represents $S_{fi}^{\mathcal{P}\gamma_D}$, $S_{fi}^{\mathcal{P}a}$, $S_{fi}^{\mu^+\gamma_D}$ or $S_{fi}^{\mu^+ a}$ ($\calm^{\mathcal{P}\gamma_D}$, $\calm^{\mathcal{P}a}$, $\calm^{\mu^+\gamma_D}$ or $\calm^{\mu^+ a}$), $\rho_\omega={a^2 \omega\over 4\pi}$ is the laser photon density~\cite{Greiner:1992bv}, and $d\Pi_2$ is the standard two-body phase space
\bee
d\Pi_2
=
\int \frac{d^3 q'}{(2 \pi)^3 2 q^{\prime 0} }
\int \frac{d^3 k'}{(2 \pi)^3 2 k^{\prime 0} }
(2\pi)^4 \delta^4( k' + q' - q - nk) \;.
\ene
In the c.m. frame, the cross section is
\begin{eqnarray}
\sigma &=&{\kappa^2 \over 32\pi q^0 \rho_\omega}\sum_{n=-\infty}^{\infty} \sum_{i,j=0}^{2}
\widetilde{C}_i^{n} \big( \widetilde{C}_j^{n} \big)^\ast  \overline{ \calm_i \calm_j^\dag }  
\int  du {1\over u^2}\;,
\end{eqnarray}
where
\begin{eqnarray}
{\sqrt{s_n}\over Q'+|\vec{q}'|}<u<{\sqrt{s_n}\over Q'-|\vec{q}'|}\;.
\end{eqnarray}
For the collision of green light laser with either muon or proton beam, it turns out that the condition $m^{\ast 2}_{\mu,\mathcal{P}}\gg \omega E_{\rm Lab}$ holds even for extremely high beam energy e.g., $E_{\rm Lab}=10$ TeV. As a result, we have the approximations as $s_n\approx m^{\ast 2}_{\mu,\mathcal{P}}$ and $u\approx 1$. Under these approximations, the squared amplitude in Eq.~(\ref{eq:amp2DP}) demonstrates that the cross section for DP production remains independent of the beam energy. However, for ALP production, the second term in the squared amplitude Eq.~(\ref{eq:amp2ALP}) is proportional to $(1-u)^2$. The magnitude by which $u$ deviates from 1 due to the beam energy change greatly influences the result. Thus, this dependence reveals that its cross section grows with the beam energy.

\begin{figure}[h]
\centering
\includegraphics[width=0.475\textwidth]{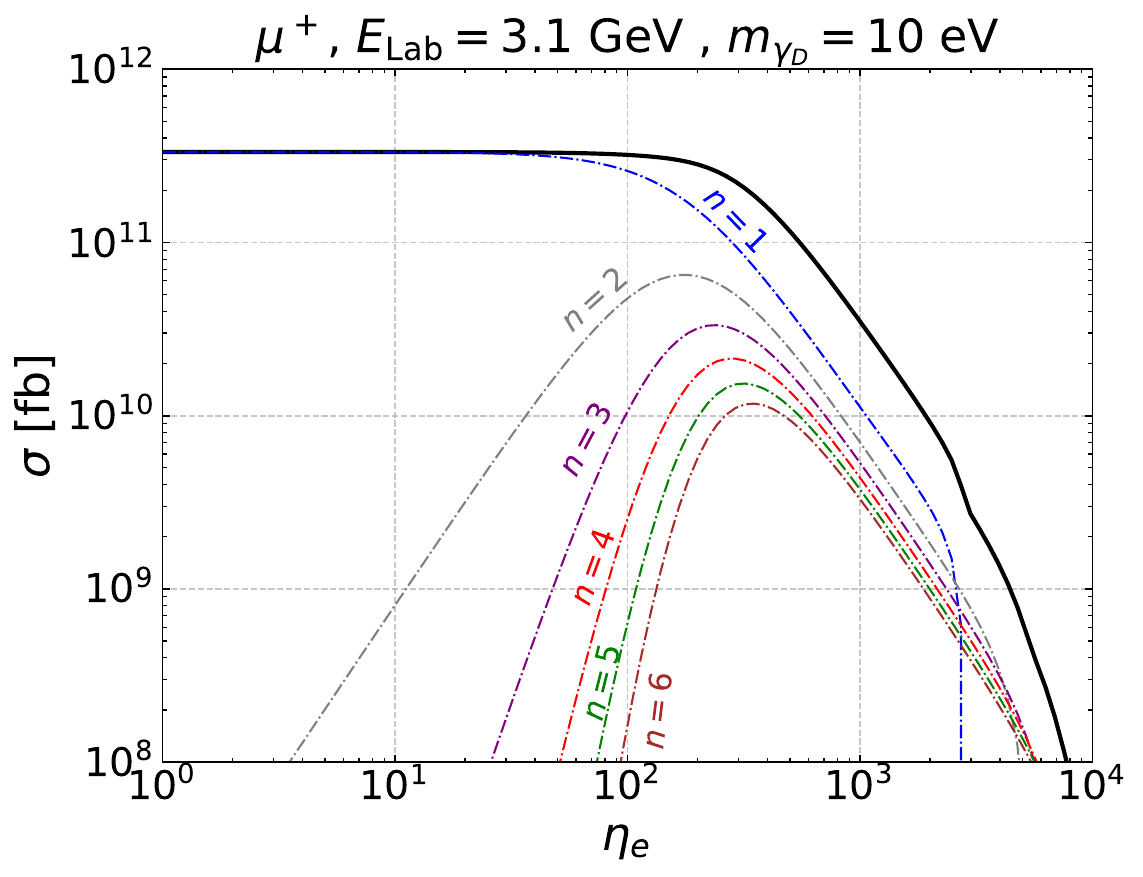}
\includegraphics[width=0.475\textwidth]{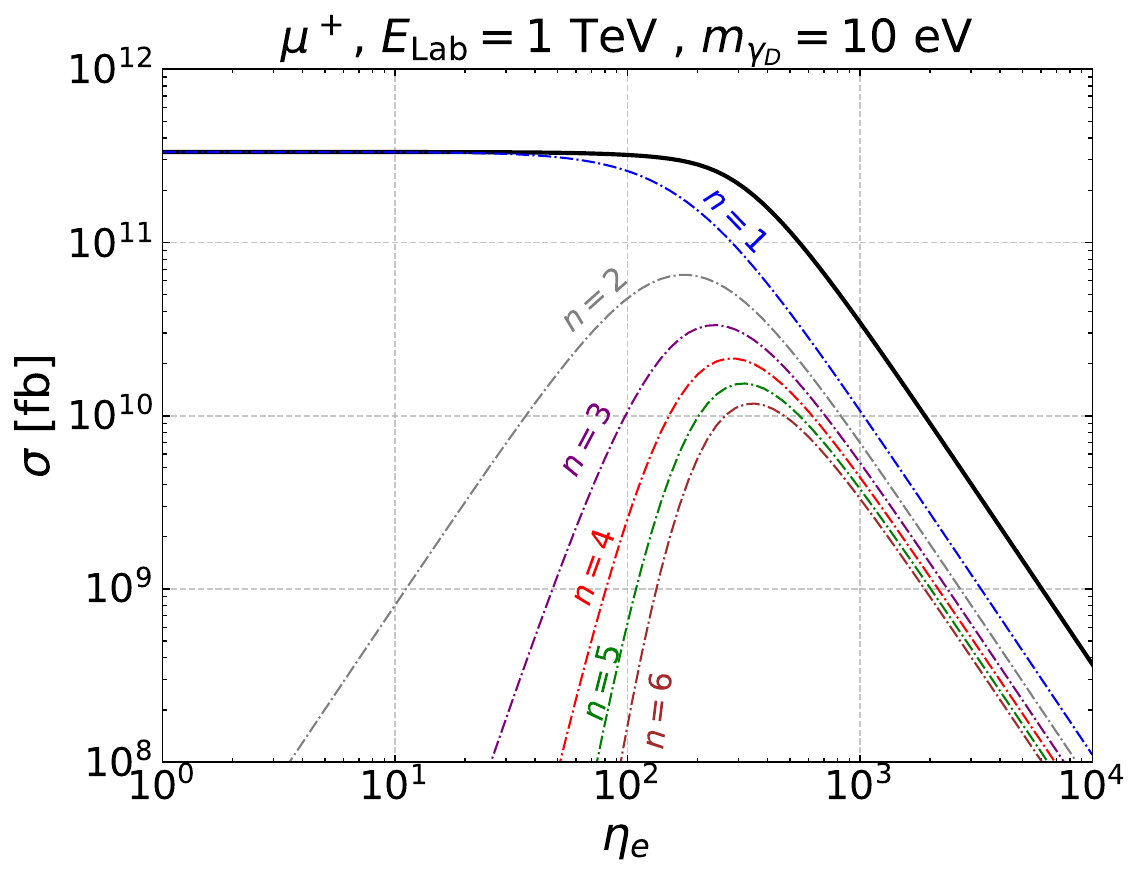}\\
\includegraphics[width=0.475\textwidth]{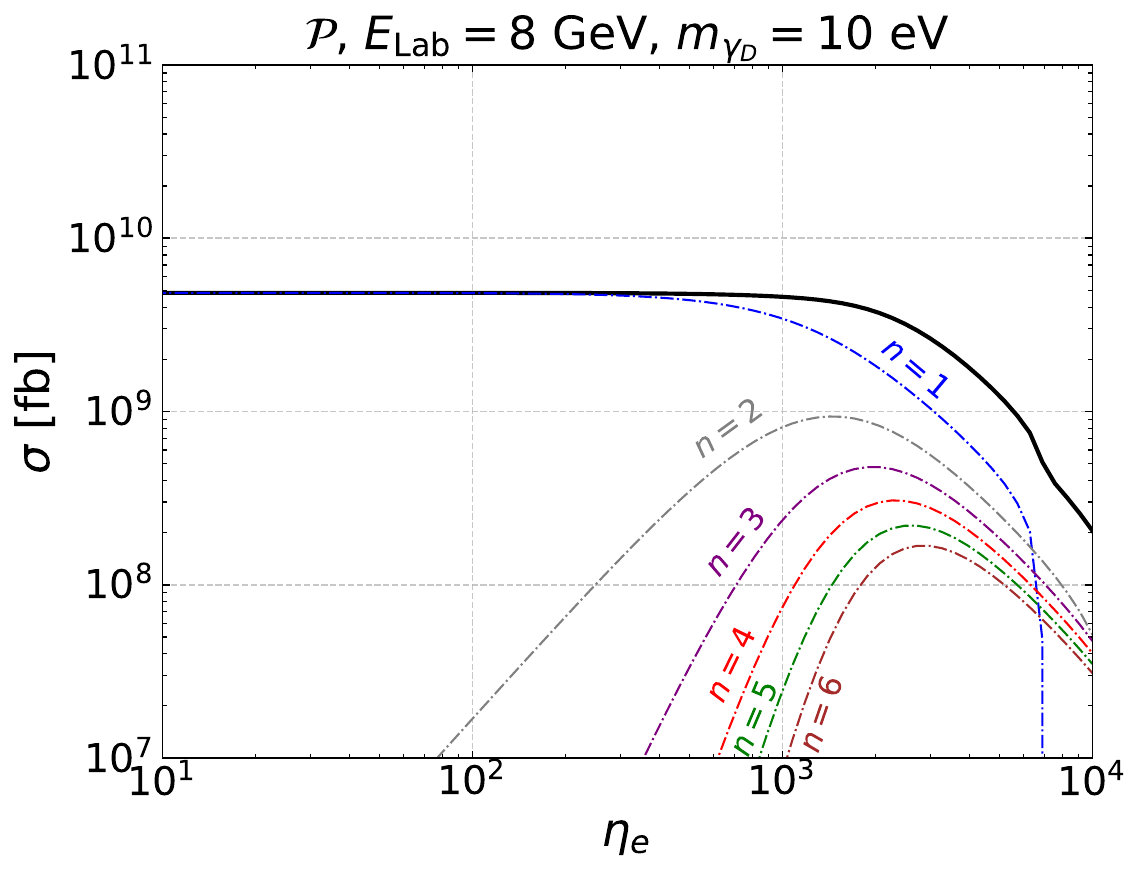}
\includegraphics[width=0.475\textwidth]{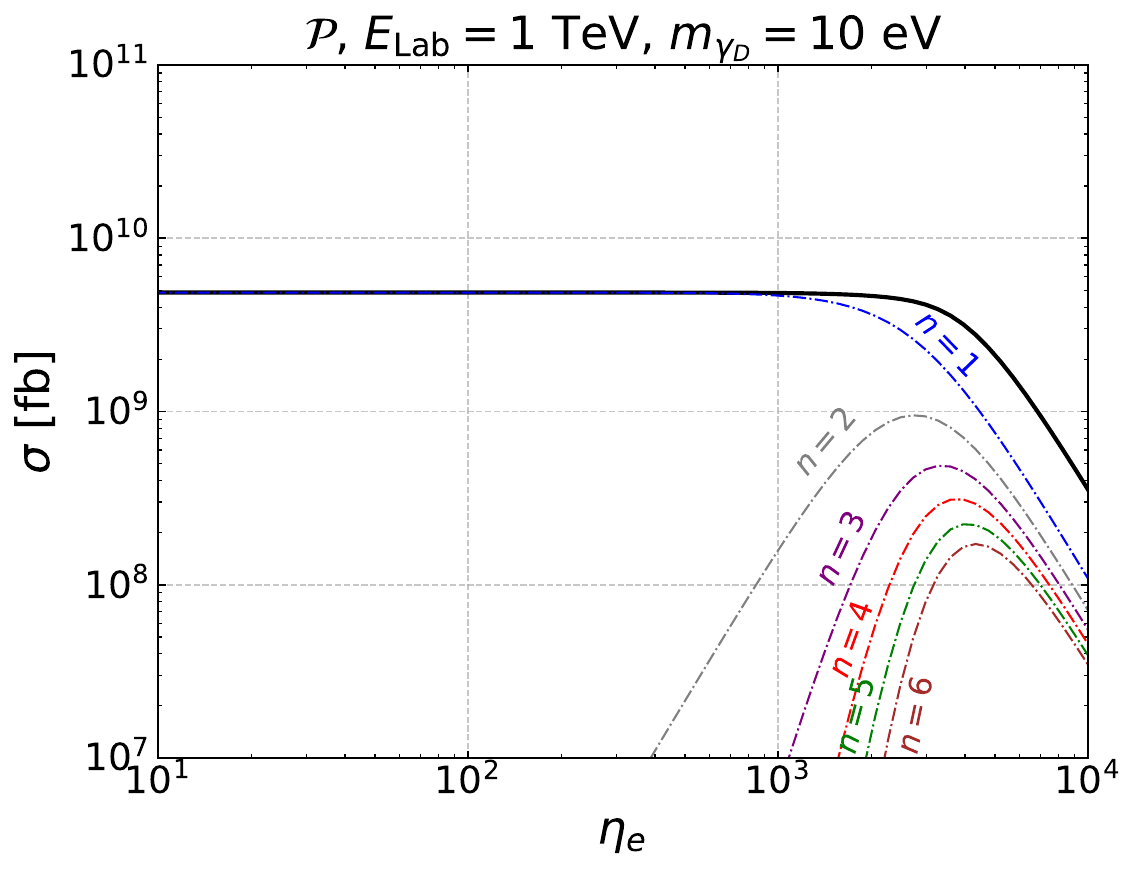}
\caption{
The cross sections for the laser-induced Compton scattering to DP as a function of $\eta_e$ with kinetic mixing $\epsilon = 1$ and DP mass $m_{\gamma_D}=10~{\rm eV}$.
The individual contributions from the absorption of $n = 1$ (blue dashed line), $n = 2$ (gray dashed line), $n=3$ (purple dashed line), $n=4$ (red dashed line), $n=5$ (green dashed line) or $n=6$ (brown dashed line) laser photons are displayed, with the totally summed cross section shown as a black solid line.
The top panels show the results of anti-muon beam with $E_{\rm Lab}=3.1~{\rm GeV}$ (left) and 1 TeV (right).
The results of the proton beam are showed in bottom panels with $E_{\rm Lab}=8~{\rm GeV}$ (left) and 1 TeV (right).
}
\label{fig:xsDP}
\end{figure}

\begin{figure}[h]
\centering
\includegraphics[width=0.475\textwidth]{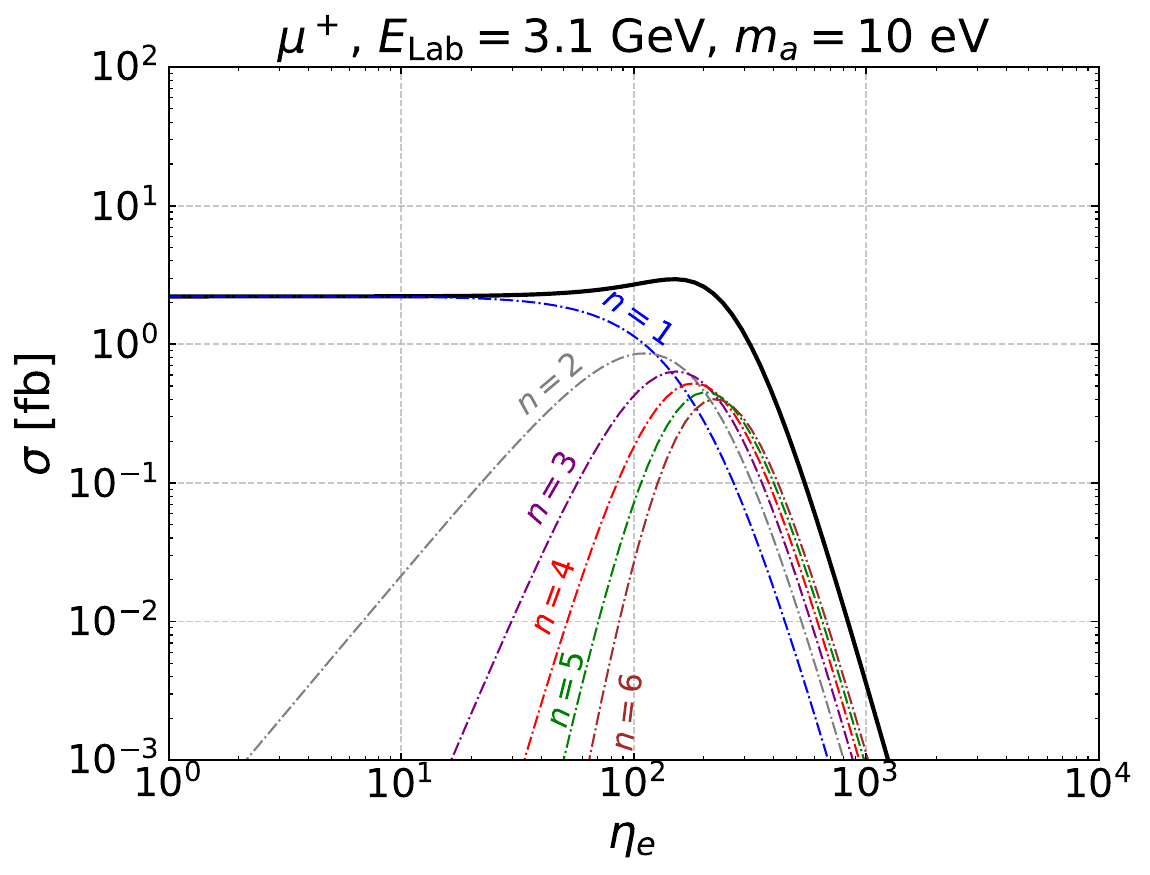}
\includegraphics[width=0.475\textwidth]{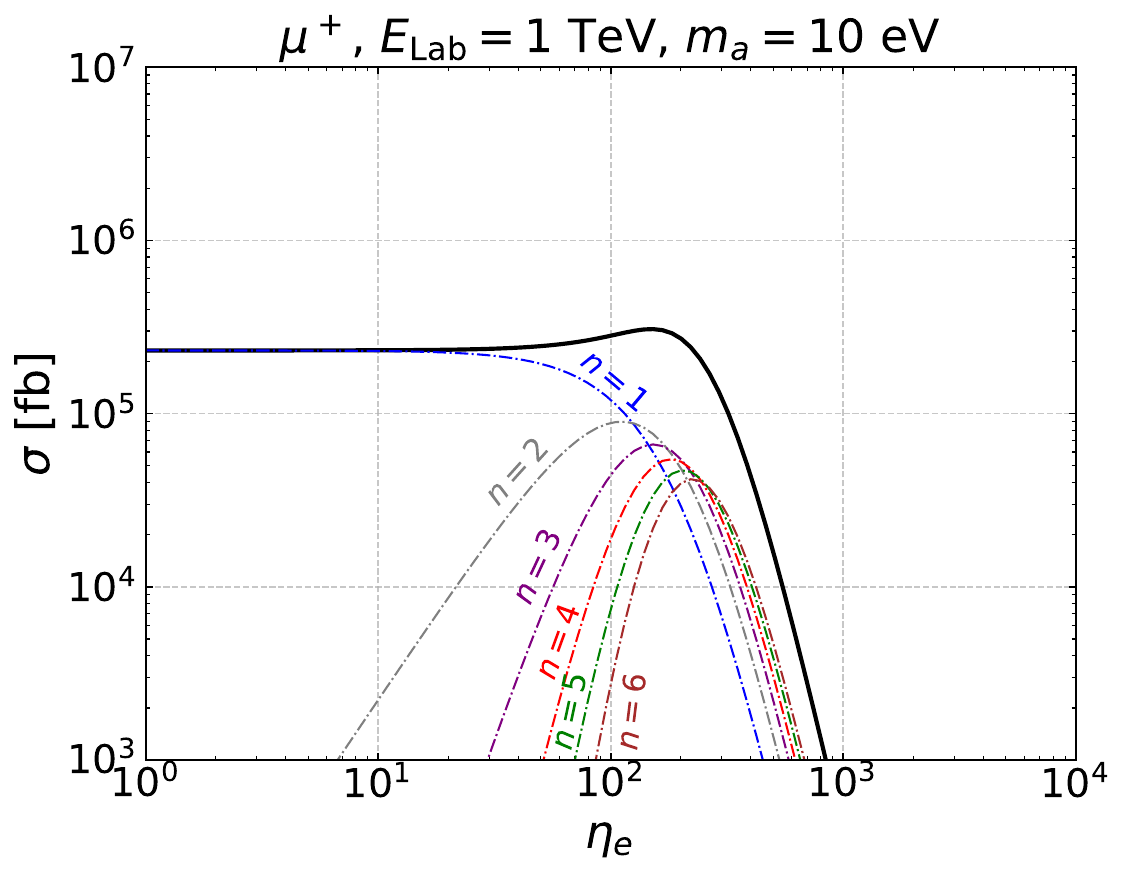}\\
\includegraphics[width=0.475\textwidth]{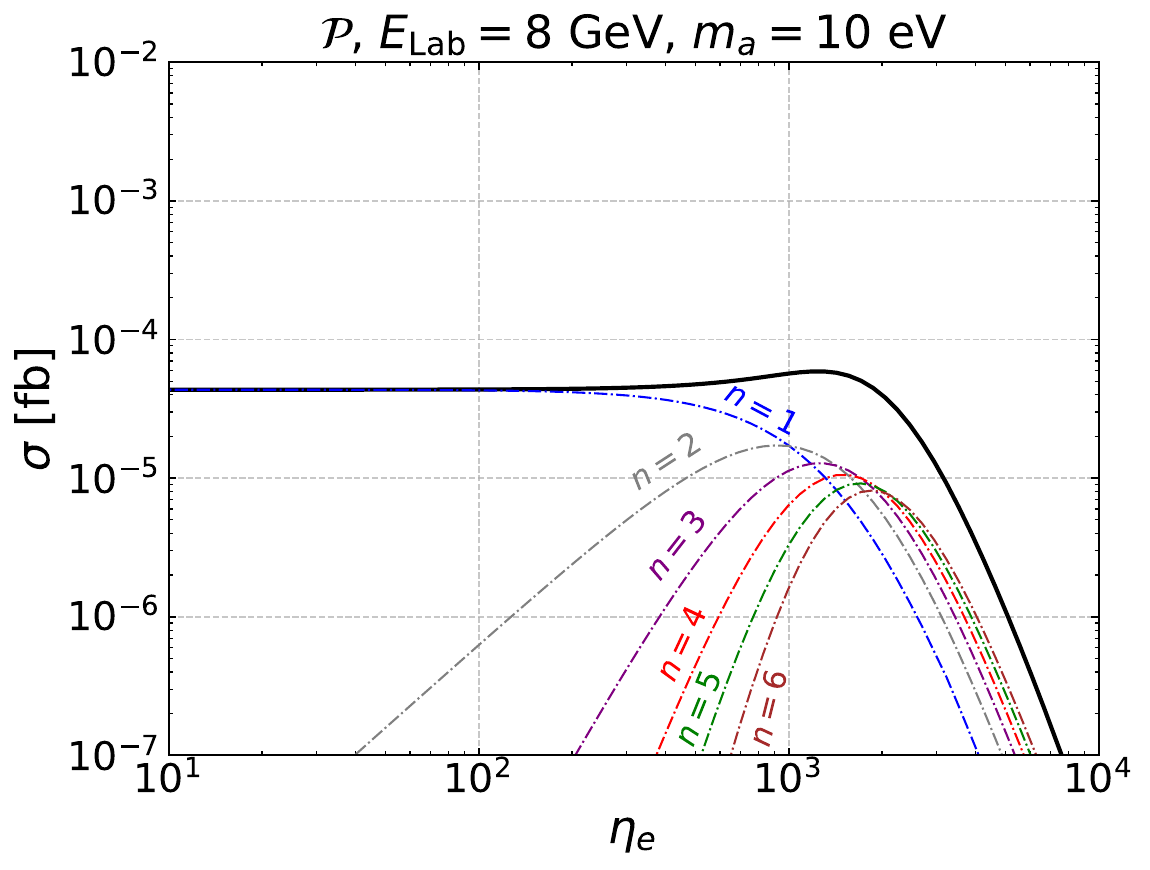}
\includegraphics[width=0.475\textwidth]{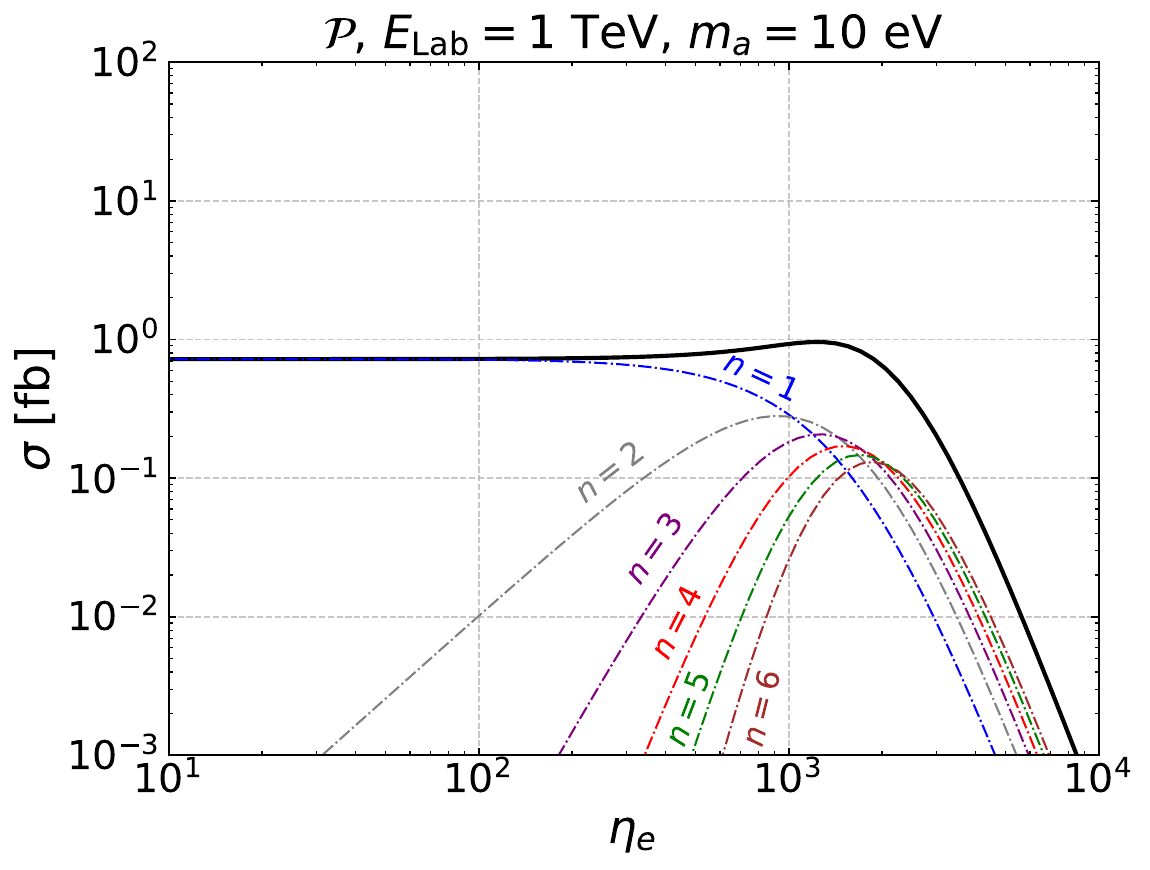}
\caption{
The cross sections for the laser-induced Compton scattering to ALP as a function of $\eta_e$ with ALP-fermion couplings $g_{a\mu}= g_{a\mathcal{P}} = 1$ and ALP mass $m_a=10~{\rm eV}$, as labeled in Fig.~\ref{fig:xsDP}.
}
\label{fig:xsALP}
\end{figure}

In Figs.~\ref{fig:xsDP} and \ref{fig:xsALP}, we show the cross sections for the laser-induced Compton scattering to DP and ALP as a function of $\eta_e$, respectively.
The individual contributions from the absorption of $n = 1-6$ laser photons are displayed, with the totally summed cross section shown as a black solid line. The top panels show the results of proton beam with $E_{\rm Lab}=8~{\rm GeV}$ (left) and 1 TeV (right). The results of the anti-muon beam are showed in bottom panels with $E_{\rm Lab}=3.1~{\rm GeV}$ (left) and 1 TeV (right). The collision of multiple laser photons results in a series of edges for the cross section beyond the exact $n=1$ cross section. For proton (muon) beam, when $\eta_e$ exceeds $10^3$ ($10^2$) and gets larger, more nonlinear effects contribute to the total cross section.
In Fig.~\ref{fig:xsDP}, one can see that the numerical cross section of DP production indeed does not change with beam energy.
Fig.~\ref{fig:xsALP} indicates that the cross section of ALP production increases with beam energy.
Note that in the calculation of exclusion bounds below, we will not consider higher values of $\eta_e$.

\subsection{Laser-induced SM backgrounds}

Next, we analyze the relevant SM background processes mediated by $W^\pm$ or $Z$ bosons in details. They give the same single fermion final state with missing neutrinos
\begin{eqnarray}
\mu^+/\mathcal{P}(p_1) + n\omega (k) \to \mu^+/\mathcal{P}(p_2) + \nu(p_4) + \overline{\nu}(p_3)\;,
\end{eqnarray}
where $p_3+p_4=k'$.

The charged and neutral weak currents in the SM result in four-fermion effective operators after integrating out the heavy $W^\pm$ and $Z$ bosons.
The relevant four-fermion Lagrangians for neutrino-muon and neutrino-quark interactions are
\begin{eqnarray}
-\mathcal{L}^{\nu \mu}&=&{G_F\over \sqrt{2}} \overline{\nu}_\ell\gamma_\mu (1-\gamma_5)\nu_\ell \overline{\mu}\gamma^\mu(g_{LV}^{\nu_\ell \mu}-g_{LA}^{\nu_\ell \mu} \gamma_5)\mu\;,\\
-\mathcal{L}^{\nu q}&=&{G_F\over \sqrt{2}} \overline{\nu}_\ell\gamma_\mu (1-\gamma_5)\nu_\ell \sum_{q=u,d,s} \Big[g_{LL}^{\nu_\ell q}\overline{q}\gamma^\mu (1-\gamma_5) q +g_{LR}^{\nu_\ell q}\overline{q}\gamma^\mu (1+\gamma_5) q  \Big]\;,
\end{eqnarray}
where
\begin{eqnarray}
g_{LV}^{\nu_\mu \mu} &=& {1\over 2} + 2\sin^2\theta_W\;,\\
g_{LA}^{\nu_\mu \mu} &=& {1\over 2}\;,\\
g_{LV}^{\nu_\ell \mu} &=& -{1\over 2} + 2\sin^2\theta_W\;,~~~\nu_\ell=\nu_e, \nu_\tau\\
g_{LA}^{\nu_\ell \mu} &=& -{1\over 2}\;,~~~\nu_\ell=\nu_e, \nu_\tau
\end{eqnarray}
and
\begin{eqnarray}
g_{LL}^{\nu_\ell u}&=&{1\over 2} - {2\over 3}\sin^2\theta_W\;,\\
g_{LL}^{\nu_\ell d,\nu_\ell s}&=&-{1\over 2} + {1\over 3}\sin^2\theta_W\;,\\
g_{LR}^{\nu_\ell u}&=&- {2\over 3}\sin^2\theta_W\;,\\
g_{LR}^{\nu_\ell d,\nu_\ell s}&=& {1\over 3}\sin^2\theta_W\;.
\end{eqnarray}
The above quark-level currents should be matched onto interactions at nucleon-level in order to calculate the laser-proton scattering cross section. We use the following relevant proton-level currents~\cite{Bishara:2017pfq}
\begin{eqnarray}
\langle \mathcal{P}(k_2) | \overline{q} \gamma^\mu q| \mathcal{P} (k_1)\rangle &=&\overline{u}_{\mathcal{P}} (k_2)\Big[F_1^{q/\mathcal{P}}(q^2)\gamma^\mu+{i\over 2m_{\mathcal{P}}}F_2^{q/\mathcal{P}}(q^2)\sigma^{\mu\nu}q_\nu\Big] u_{\mathcal{P}}(k_1)\;,
\label{eq:protonV}\\
\langle \mathcal{P}(k_2) | \overline{q} \gamma^\mu \gamma_5 q| \mathcal{P} (k_1)\rangle &=&\overline{u}_{\mathcal{P}} (k_2)\Big[F_A^{q/\mathcal{P}}(q^2)\gamma^\mu \gamma_5+{1\over 2m_{\mathcal{P}}}F_{P'}^{q/\mathcal{P}}(q^2)\gamma_5 q^\mu\Big] u_{\mathcal{P}}(k_1)\;,
\label{eq:protonAV}
\end{eqnarray}
where $q^\mu=k_2^\mu - k_1^\mu$. As the transfer momentum squared $q^2$ is very small in our case, we adopt the form factors for the proton in the limit of $q^2=0$~\cite{Bishara:2017pfq}
\begin{eqnarray}
&&F_1^{u/\mathcal{P}}(0)=2\;,~~~F_1^{d/\mathcal{P}}(0)=1\;,~~~F_1^{s/\mathcal{P}}(0)=0\;,\\
&&F_2^{u/\mathcal{P}}(0)=1.609(17)\;,~~~F_2^{d/\mathcal{P}}(0)=-2.097(17)\;,~~~F_2^{s/\mathcal{P}}(0)=-0.064(17)\;,\\
&&F_A^{u/\mathcal{P}}(0)=\Delta u_\mathcal{P}\;,~~~F_A^{d/\mathcal{P}}(0)=\Delta d_\mathcal{P}\;,~~~F_A^{s/\mathcal{P}}(0)=\Delta s_\mathcal{P}\;,\\
&&F_{P'}^{q/\mathcal{P}}(0)={m_\mathcal{P}^2\over m_\pi^2} a_{P',\pi}^{q/\mathcal{P}}+{m_\mathcal{P}^2\over m_\eta^2} a_{P',\eta}^{q/\mathcal{P}}+b_{P'}^{q/\mathcal{P}}\;,~~~q=u,d,s\;,
\end{eqnarray}
where $a_{P',\pi}^{u/\mathcal{P}}=-a_{P',\pi}^{d/\mathcal{P}}=2(\Delta u_\mathcal{P}-\Delta d_\mathcal{P})$, $a_{P',\pi}^{s/\mathcal{P}}=0$, $a_{P',\eta}^{u/\mathcal{P}}=a_{P',\eta}^{d/\mathcal{P}}=-a_{P',\eta}^{s/\mathcal{P}}/2=2(\Delta u_\mathcal{P}+\Delta d_\mathcal{P}-2\Delta s_\mathcal{P})/3$, $\Delta u_\mathcal{P}=0.897(27)$, $\Delta d_\mathcal{P}=-0.376(27)$, $\Delta s_\mathcal{P}=-0.031(5)$, $b_{P'}^{u/\mathcal{P}}\approx -4.65(25)$, $b_{P'}^{d/\mathcal{P}}\approx 3.28(25)$, $b_{P'}^{s/\mathcal{P}}\approx 0.32(18)$. Next, we only keep the leading-order terms in Eqs.~\eqref{eq:protonV} and \eqref{eq:protonAV} (vector and axial-vector currents) in our calculation.

The effective Lagrangian can be given as
\begin{eqnarray}
-\mathcal{L}^{\nu f}
&\approx& \frac{G_{F}}{\sqrt{2}}\Big(\bar{f}~\gamma_\mu(\Gamma_V^f -\Gamma_A^f\gamma_5)~ f\Big)\Big(\bar{\nu}_\ell~\gamma^{\mu}(1-\gamma^{5})\nu_\ell \Big)\;,~~f=\mathcal{P}, \mu\;,
\end{eqnarray}
where $\Gamma_V^{\mathcal{P}}\equiv \sum_{q=u,d,s} (g_{LL}^{\nu_\ell q}+ g_{LR}^{\nu_\ell q})~F_{1}^{q/\mathcal{P}}(0)$, $\Gamma_A^{\mathcal{P}}\equiv \sum_{q=u,d,s}(g_{LL}^{\nu_\ell q}- g_{LR}^{\nu_\ell q})~F_{A}^{q/\mathcal{P}}(0)$, $\Gamma_V^{\mu}=g_{LV}^{\nu_\ell \mu}$, and $\Gamma_A^{\mu}=g_{LA}^{\nu_\ell \mu}$.
The $S$ matrix element for the production of neutrinos from the collision between laser and proton beam becomes
\begin{eqnarray}
S_{fi}^{\nu \mathcal{P}} &=&-i\frac{G_F}{\sqrt{2}}\int d^4x  \Big(\overline{\psi}_{p_2, s_2}(x) \gamma_\mu(\Gamma_V^\mathcal{P} -\Gamma_A^\mathcal{P}\gamma_5) \psi_{p_1, s_1}(x)\Big)  \Big(\overline{\psi}_{p_4, s_4}(x) \gamma^\mu(1-\gamma_5)  \psi_{p_3, s_3}(x)\Big) \nonumber\\
&=& -i\frac{G_F}{\sqrt{2}} \int d^4 x~ e^{-i(q_1-q_2-p_3-p_4) \cdot x} \mathcal{M}^{\nu \mathcal{P}}~e^{-i\Phi}\;,
\end{eqnarray}
where the wave functions of dressed proton and free neutrinos are
\begin{eqnarray}
\psi_{p_1, s_1}(x) & = & \left[ 1 + \frac{e\cancel{k}\cancel{A}}{2k\cdot p_1} \right] u(p_1,s_1) ~e^{iF(q_1,s_1)}\;, \\[0.5em]
\overline{\psi}_{p_2, s_2}(x) & = & \overline{u}(p_2, s_2)  ~e^{-i F(q_2, s_2)} \left[ 1 + \frac{e\cancel{A}\cancel{k}}{2k\cdot p_2} \right]\;, \\[0.5em]
\psi_{p_3, s_3}(x) & = & v(p_3,s_3) ~e^{ip_3 \cdot x}\;, \\[0.5em]
\overline{\psi}_{p_4, s_4}(x) & = & \overline{u}(p_4, s_4) ~e^{i p_4 \cdot x}\;.
\end{eqnarray}
For this calculation, we use the method of helicity amplitude in terms of the density matrices of production and decay parts.
The total amplitude can be written as a product of two currents
\begin{eqnarray}
\mathcal{M}^{\nu\mathcal{P}} &=& \Big[ \bar{u}(p_2,s_2)(1 + \frac{e\cancel{A}\cancel{k}}{2k\cdot p_2})\gamma_\mu(\Gamma_V^{\mathcal{P}}-\Gamma_A^{\mathcal{P}}\gamma_5)( 1 + \frac{e\cancel{k}\cancel{A}}{2k\cdot p_1})u(p_1,s_1)\Big]\Big[ \bar{u}(p_4,s_4)\gamma^\mu(1-\gamma_5)u(p_3,s_3)\Big] \nonumber \\[1em]
&=& g^{\mu\nu}\mathcal{M}_{P,\mu}^{\nu\mathcal{P}} \mathcal{M}_{D,\nu}^{\nu\mathcal{P}}\;,
\end{eqnarray}
where the amplitudes $\mathcal{M}_{P}^{\nu\mathcal{P}}$ and $\mathcal{M}_{D}^{\nu\mathcal{P}}$ describe the production and decay of a fictitious spin-one particle
with momentum $k'$, respectively, and both of them are Lorentz invariant.
After employing the helicity amplitude method, this amplitude can be decomposed into two Lorentz invariant amplitudes along the momentum $k'$
\begin{eqnarray}
\mathcal{M}^{\nu\mathcal{P}} &=&\sum_{\lambda=s,0,\pm 1} \eta_\lambda \mathcal{M}_{P,\lambda}^{\nu\mathcal{P}} \mathcal{M}_{D,\lambda}^{\nu\mathcal{P}}\;,
\end{eqnarray}
where $\lambda=s,0,\pm 1$ is the helicity projected
along the momentum $k'$, and $\mathcal{M}_{P,\lambda}^{\nu\mathcal{P}}=\varepsilon_\lambda^\ast \cdot \mathcal{M}_{P}^{\nu\mathcal{P}}$ and $\mathcal{M}_{D,\lambda}^{\nu\mathcal{P}}=\varepsilon_\lambda \cdot \mathcal{M}_{D}^{\nu\mathcal{P}}$ with $\varepsilon_\lambda^{(\ast)\mu}$ being the polarization vectors.

By considering the absorption of $n$ laser photons, the $S$ matrix can be rewritten as
\begin{eqnarray}
S_{fi}^{\nu\mathcal{P}}&=&-i\frac{G_F}{\sqrt{2}} \sum_{n=-\infty}^{\infty} \int d^4 x~ e^{-i(q_1+nk-q_2-p_3-p_4) \cdot x} \mathcal{M}^{\nu \mathcal{P}}_n(z)\;,
\end{eqnarray}
where the $n$-th amplitude is $\mathcal{M}^{\nu \mathcal{P}}_n(z)=\mathcal{M}^{\nu \mathcal{P}}_{P,n}(z)\cdot \mathcal{M}^{\nu \mathcal{P}}_{D}$ with $\mathcal{M}^{\nu \mathcal{P}}_{P,n}(z)=B_n(z)\mathcal{M}^{\nu \mathcal{P}}_{P}$. Then, the total cross section becomes
\begin{eqnarray}
\sigma^{\nu\mathcal{P}}&=&{1\over 2\rho_\omega} \frac{G_F^2}{2} \sum_{n=-\infty}^{\infty} \frac{1}{2q^0_1} \int d\Pi_{P,n} \int \frac{dm_{k'}^2}{2\pi} \sum_{\lambda,\lambda'=s,0,\pm 1} \eta_\lambda \eta_{\lambda'} \mathcal{P}_{n,\lambda\lambda'}^{\nu\mathcal{P}} \overline{\mathcal{D}_{\lambda\lambda'}^{\nu\mathcal{P}}}\;,
\end{eqnarray}
where
\begin{eqnarray}
d\Pi_{P,n} = \frac{1}{16\pi^{2}u^{2}}  du  d\phi_{n}^{*}
\end{eqnarray}
with $\phi_{n}^{*}$ being the azimuthal angle of the momentum $k'$. After integrating out the two-body phase space
of the outgoing SM neutrinos, decay density matrices can be easily calculated. The non-zero density matrices of decay and production parts are
\begin{eqnarray}
\overline{\mathcal{D}_{ss}^{\nu\mathcal{P}}}  &=& \displaystyle\frac{m_{k^\prime}^2}{4\pi}  \beta \left( 1 -  \beta^{2} \right) \;,~~\beta=\sqrt{1-4m_\nu^2/m_{k'}^2}\approx 1\nonumber \\[0.5em]
\overline{\mathcal{D}_{\lambda\lambda}^{\nu\mathcal{P}}} & =& \displaystyle\frac{m_{k^\prime}^2}{4\pi}  \beta \left( 1 + \frac{1}{3} \beta^{2} \right), ~~~~\text{for } \lambda = 0, \pm 1
\end{eqnarray}
and
\begin{eqnarray}
\mathcal{P}_{n,ss}^{\nu\mathcal{P}} & = & 4J_n^2~(\Gamma_A^{\mathcal{P}})^2m_\mathcal{P}^2 +4e^2a^2\mathcal{J}_n~ (\Gamma_A^{\mathcal{P}})^2\frac{m_\mathcal{P}^2}{m_{k^\prime}^2}u(1-\frac{1}{u})^2\;, \nonumber \\[0.5em]
\sum_{\lambda=0,\pm1} \mathcal{P}_{n,\lambda\lambda}^{\nu\mathcal{P}} & = & -2 J_n^2~\Big[(\Gamma_V^\mathcal{P})^2(m_{k^\prime}^2+2m_\mathcal{P}^2)+(\Gamma_A^\mathcal{P})^2(m_{k^\prime}^2-4m_\mathcal{P}^2) \Big] \nonumber \\
&& - 4e^2a^2\mathcal{J}_n ~\Big[ (\Gamma_V^\mathcal{P})^2(\frac{u}{2}+\frac{2}{u})+(\Gamma_A^\mathcal{P})^2\Big( \frac{1}{2}(u+\frac{1}{u})+\frac{m_\mathcal{P}^2}{m_{k^\prime}^2}u(1-\frac{1}{u})^2   \Big) \Big]\;, \nonumber
\end{eqnarray}
where $\mathcal{J}_n\equiv J_n^2-{1\over 2}\Big(J_{n-1}^2+J_{n+1}^2\Big)$. For laser-muon scattering, we can obtain relevant results with substitutions $m_\mathcal{P}\to m_\mu$, $\Gamma_V^\mathcal{P}\to \Gamma_V^\mu$ and $\Gamma_A^\mathcal{P}\to \Gamma_A^\mu$ in the above formulas. In Table~\ref{tab:xsecbkg}, we show the numerical cross sections of background $\mu^+/\mathcal{P}+{\rm Laser}\to \mu^+/\mathcal{P}+\nu +\overline{\nu}$. They are negligibly small compared with our signal processes.
Nevertheless, we take into account this background in the following analysis of prospective sensitivity.

\begin{table}[htbp]
\centering
\begin{tabular}{l|c|c|c|c}
\hline\hline
$\sigma_{\rm bkg}$ [fb] & $\mu^+(\eta_e=0.1)$ & $\mu^+(\eta_e=100)$ & $\mathcal{P}(\eta_e=0.1)$  & $\mathcal{P}(\eta_e=100)$ \\
\hline\hline
$E_{\rm Lab}$=3.1 GeV for $\mu^+$ & 3.12$\times 10^{-28}$ & $1.12\times 10^{-27}$ & $1.09\times 10^{-31}$ & $1.12\times 10^{-31}$\\
~~~~~~~~~~~8 GeV for $\mathcal{P}$ &&&&\\
\hline
$E_{\rm Lab}$=1 TeV  & $1.10\times 10^{-15}$ & $1.12\times 10^{-15}$ & $3.42\times 10^{-21}$ & $3.53\times 10^{-21}$\\
\hline
$E_{\rm Lab}$=10 TeV & $1.08\times 10^{-10}$ & $3.80\times 10^{-10}$ & $3.42\times 10^{-16}$  & $3.53\times 10^{-16}$ \\
\hline\hline
\end{tabular}
\caption{The cross sections of background $\mu^+/\mathcal{P}+{\rm Laser}\to \mu^+/\mathcal{P}+\nu +\overline{\nu}$. We take $\eta_e=0.1$ or 100 and $E_{\rm Lab}=3.1~{\rm GeV}$ (for muon beam) or 8 GeV (for proton beam), 1 TeV and 10 TeV.}
\label{tab:xsecbkg}
\end{table}

\section{Prospective sensitivity}
\label{sec:results}

In this section, we show the prospective sensitivity of Compton scattering to DP or ALP coupling through the collision between laser pulse and muon or proton beam.
The muon (proton) beam is supposed to have the energy of $E_{\rm Lab}=3.1$ GeV, 1 TeV or 10 TeV ($E_{\rm Lab}=8$ GeV, 1 TeV or 10 TeV). It collides with
an intense laser beam of green light with $\omega=2.35$ eV~\cite{Bamber:1999zt,LUXE:2023crk} and the initial scattering angle as $\theta_{\rm Lab}=17^\circ$~\cite{Bamber:1999zt,LUXE:2023crk}.
The dark particles are assumed to be lighter than $2m_e\approx 1~{\rm MeV}$. They are thus long-lived and become invisible in the experiment.
We thus propose to search for the signal of single muon or proton with missing energy. Although this signal is quite common at colliders, the simulation of missing energy event production in laser experiments is beyond the scope of this theoretical work.
Given the SM background calculated above, we require a
$3\sigma$ observation based on the significance formula~\cite{ParticleDataGroup:2024cfk}
\begin{eqnarray}
{S\over \sqrt{S+B}}\;,
\end{eqnarray}
where $S$ ($B$) denotes the number of signal (background) events.
We assume integrated luminosity $\mathcal{L}=1~{\rm ab}^{-1}$ to get the sensitivity limits on dark particle couplings. A more detailed study of missing energy events in laser experiments is expected in future work.
In the calculation of significance, only the irreducible contribution from SM with missing neutrinos is considered as
the background. In principle, there can be other backgrounds from reducible processes. For instance, the photons in the nonlinear Compton scattering $\mu^+/\mathcal{P}\to \mu^+/\mathcal{P}+n\gamma$ are not recorded by
the detector. When the energy threshold of electron-positron pair is achieved, the trident electron-positron pair production $\mu^+/\mathcal{P}\to \mu^+/\mathcal{P}+e^+e^-$ can also contribute to background if the electron-positron pair is misidentified in the detector. However, the contributions of all these processes are reducible and cannot dominate over the irreducible contribution. Considering only the irreducible background is sufficient for a primary estimation of the experimental sensitivity.

\begin{figure}[ht]
\begin{center}
\includegraphics[height=0.383\textwidth]{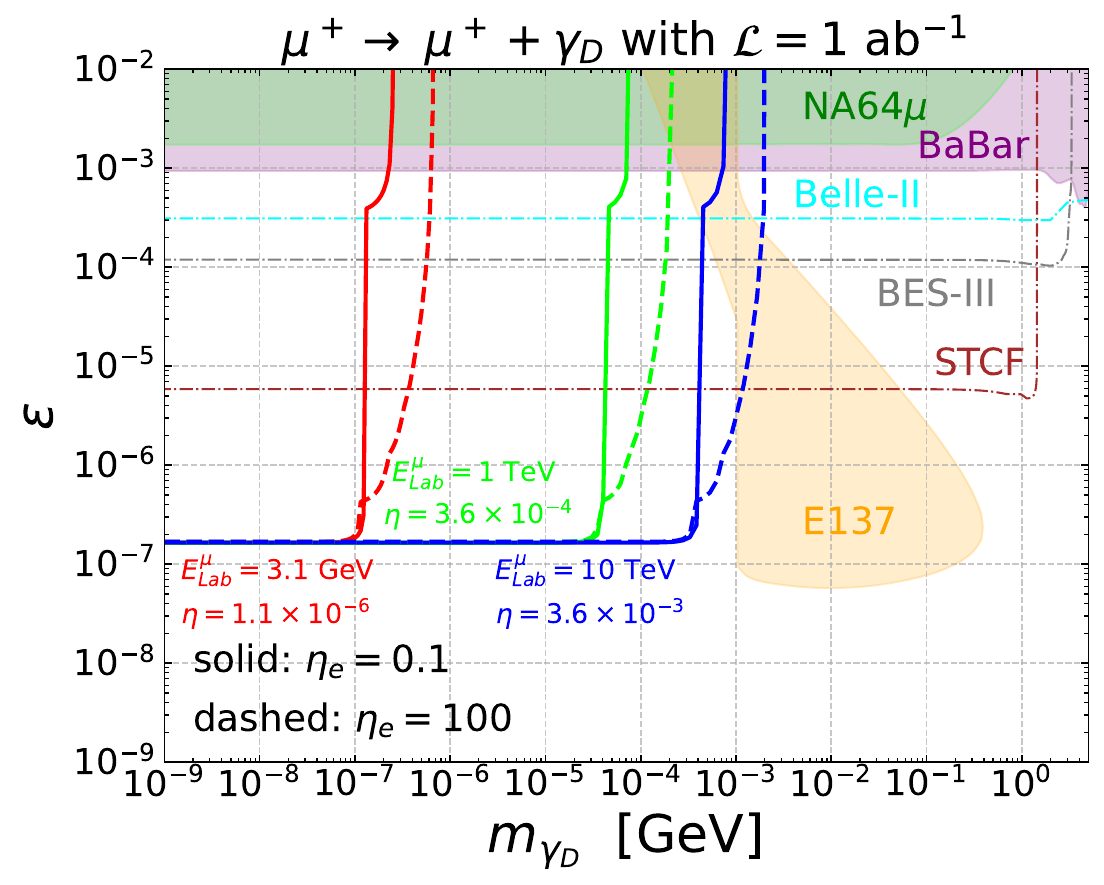}
\includegraphics[height=0.383\textwidth]{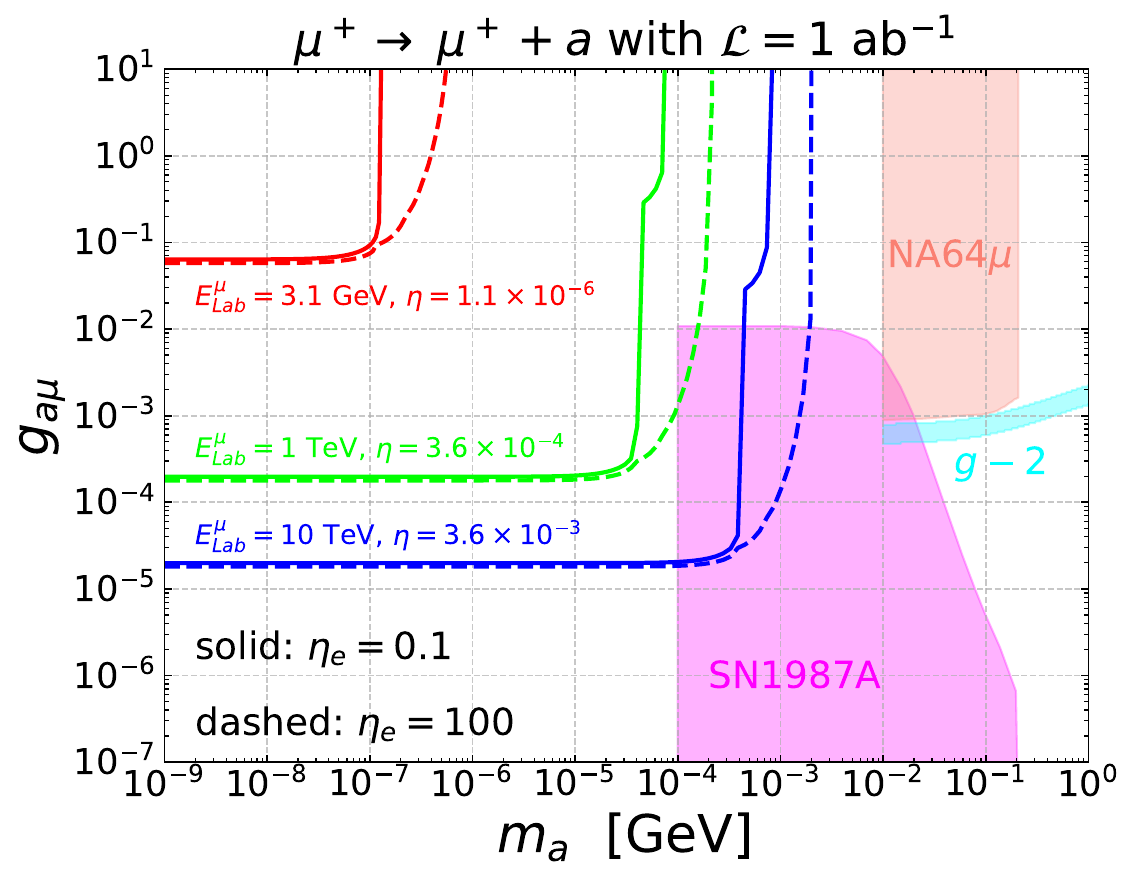}\\
\includegraphics[height=0.383\textwidth]{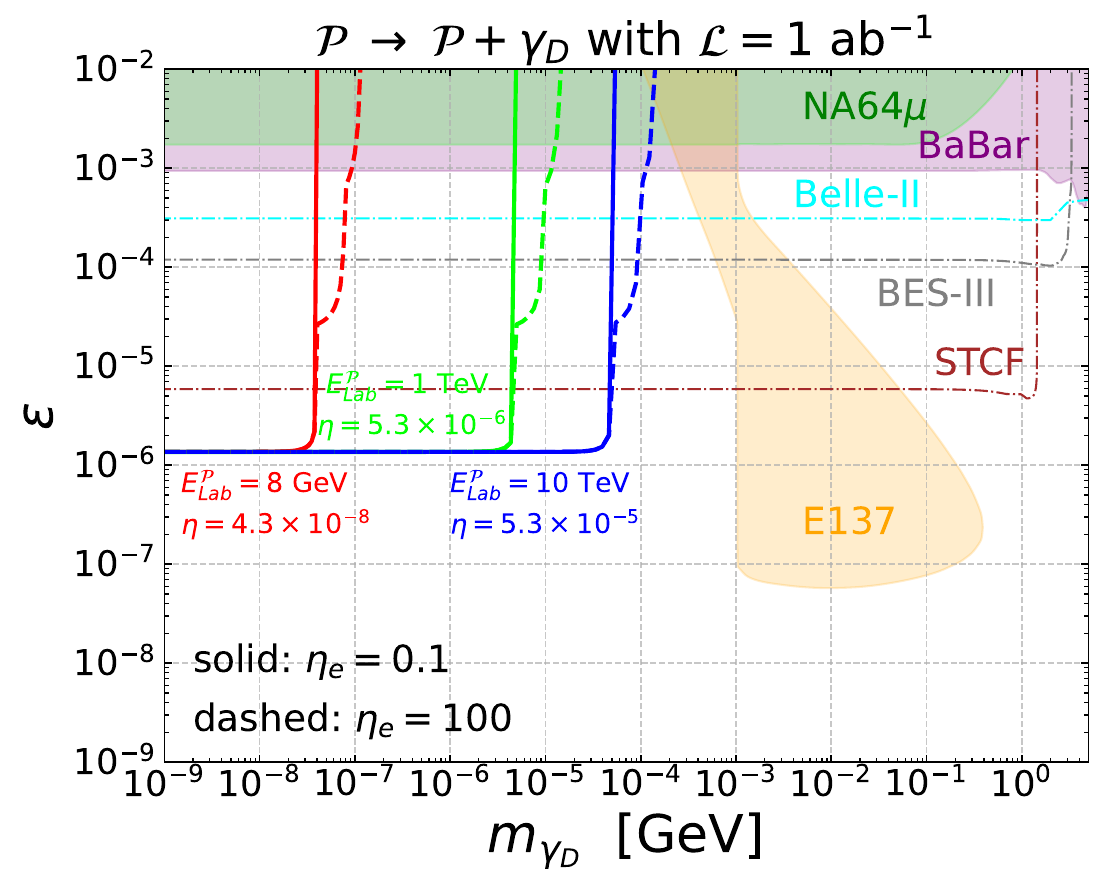}
\includegraphics[height=0.383\textwidth]{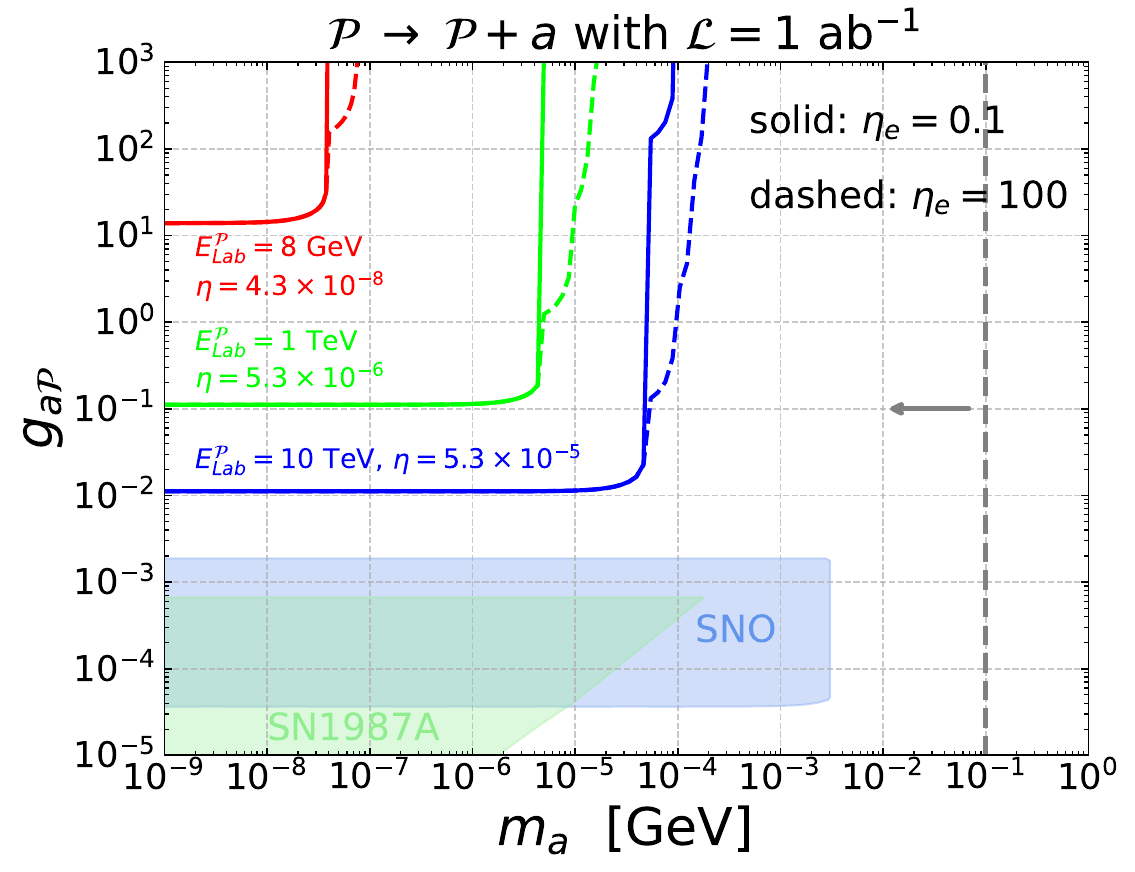}
\caption{Sensitivity of nonlinear Compton scattering to the kinetic mixing of DP $\epsilon$ (left panels) and the ALP coupling $g_{a\mu}$ or $g_{a\mathcal{P}}$ (right panels) from the collision between laser and muon (top panels) or proton (bottom panels) beam. We take $\eta_e=0.1$ (solid lines) or 100 (dashed lines) and $E_{\rm Lab}=3.1~{\rm GeV}$ or 8 GeV (red), 1 TeV (green) and 10 TeV (blue). The limits from other experiments are also shown. For DP, we include the exclusion limits from BaBar (light blue purple region)~\cite{BaBar:2017tiz}, E137 (orange region)~\cite{Liu:2017htz}, NA64$\mu$ (light green region)~\cite{Andreev:2024yft}, and the projection limits from Belle II ($20~{\rm fb}^{-1}$, cyan line)~\cite{Belle-II:2018jsg}, BESIII ($17~{\rm fb}^{-1}$, gray line)~\cite{Zhang:2019wnz} and STCF (2 GeV, $30~{\rm ab}^{-1}$, brown line)~\cite{Zhang:2019wnz}. For ALP coupling $g_{a\mu}$, we include exclusion limits from SN1987A (purple region)~\cite{Croon:2020lrf}, NA64$\mu$ ($2\times 10^{10}$ muons on target, salmon region)~\cite{Li:2025yzb} and $(g-2)_\mu$ $2\sigma$ band (cyan region)~\cite{Buen-Abad:2021fwq}. For ALP coupling $g_{a\mathcal{P}}$, we include exclusion limits from SNO (light blue region)~\cite{Bhusal:2020bvx} and SN1987A (light green region)~\cite{Lella:2023bfb}.
The region to the left of dotted line is excluded by the probe of ALP in flavor experiments~\cite{Bauer:2021mvw,Li:2025ski}.
}
\label{fig:Sen}
\end{center}
\end{figure}

Fig.~\ref{fig:Sen} shows the sensitivity of nonlinear Compton scattering to the kinetic mixing of DP $\epsilon$ (left panels) and the ALP coupling $g_{a\mu}$ or $g_{a\mathcal{P}}$ (right panels) from the collision between laser and muon (top panels) or proton (bottom panels) beam.
The limits from other experiments are also shown for comparison.
The limit of DP kinetic mixing can reach $\epsilon\sim 1.6 \times 10^{-7}$ ($1.4 \times 10^{-6}$) for muon (proton) beam and is independent of beam energy. One can see that the laser-induced process provides a complementary search of dark photon for $m_{\gamma_D}<1$ MeV, compared with other collider and beam dump experiments.

The muon beam collision can probe $g_{a\mu}$ as low as $5\times 10^{-2}$ ($2\times 10^{-4}$) [$2\times 10^{-5}$] for $E_{\rm Lab}=3.1~{\rm GeV}$ (1 TeV) [10 TeV]. The reachable parameter space is broader than that from Supernova 1987A (SN1987A)~\cite{Croon:2020lrf} for $g_{a\mu}>10^{-2}$ (see more discussions about SN1987A bound in Refs.~\cite{Caputo:2022rca,Caputo:2021rux,Fiorillo:2025yzf}). The limit of $g_{a\mu}$ for $E_{\rm Lab}=1$ TeV or 10 TeV is at least one order of magnitude lower than the expected bound from muon beam dump experiment (NA64$\mu$) with $2\times 10^{10}$ muons on target~\cite{Li:2025yzb}~\footnote{Note that for other constraints, we typically stop at the lower mass limit indicated in the figure of literature. But in fact, the limit data can extend to even lower masses.} for $m_a<1$ MeV. The sensitivity of proton beam collision to $g_{a\mathcal{P}}$ is three orders of magnitude weaker. It is complementary to the constraints from SN1981A~\cite{Lella:2023bfb} and Sudbury Neutrino Observatory (SNO)~\cite{Bhusal:2020bvx} experiments for $g_{a\mathcal{P}}>1\times 10^{-2}$.~\footnote{There exists early explanation of gamma ray bursts in terms of ALP-nucleon coupling~\cite{Berezhiani:1999qh}. This model also implies $g_{a\mu}\sim 10^{-5}-10^{-8}$.} However, the fundamental ALP-quark couplings are strongly constrained by the probe of ALP via $K\to \pi+a$ decays etc. in flavor experiments~\cite{Bauer:2021mvw,Li:2025ski} although the constraint is model-dependent. After converting the universal ALP-quark coupling bound, we find the region of $g_{a\mathcal{P}}\gtrsim 10^{-6}$ with $m_a\lesssim 0.1$ GeV is entirely excluded by flavor constraints, as shown by the region to the left of dotted line in the lower right panel of Fig.~\ref{fig:Sen}. One can see that increasing beam energy or $\eta_e$ from 0.1 to 100 allows access to higher mass region. We also show the values of another energy quantity $\eta\equiv k\cdot p/m^2$ in the plots.

\section{Conclusion}
\label{sec:Con}

The laser of an intense electromagnetic field promotes the studies of strong-field particle physics in high-intensity frontier. Nowadays, particle accelerator facilities in the world provide the high-quality muon and proton beams. They motivate the access of new physics beyond the SM by combining intense lasers and energetic muon or proton beams.

In this work, we propose the nonlinear Compton scattering to dark particles through the collision between laser pulse and muon or proton beam. We take dark photon and axion-like particle as illustrative dark particles and assume the existence of their couplings to muon and proton.
The cross sections of nonlinear Compton scattering to light dark photon or axion-like particle are calculated in strong-field QED. We also analyze the background processes with missing neutrinos. We find that
\begin{itemize}
\item The absorption of multiple laser photons results in a series of edges for the cross section beyond the exact $n=1$ result. For proton (muon) beam, when $\eta_e$ exceeds $10^3$ ($10^2$) and gets larger, more nonlinear effects contribute to the total cross section.
\item The limit of DP kinetic mixing can reach $\epsilon\sim 1.6 \times 10^{-7}$ ($1.4 \times 10^{-6}$) for muon (proton) beam and is independent of beam energy.
\item The muon beam collision can probe ALP-muon coupling $g_{a\mu}$ as low as $5\times 10^{-2}$ ($2\times 10^{-4}$) [$2\times 10^{-5}$] for $E_{\rm Lab}=3.1~{\rm GeV}$ (1 TeV) [10 TeV]. The sensitivity of proton beam collision to ALP-proton coupling $g_{a\mathcal{P}}$ is three orders of magnitude weaker.
\item The laser-induced Compton scattering provides a complementary and competitive probe of dark particles with mass less than 1 MeV, compared with other experiments. Increasing beam energy or intensity parameter $\eta_e$ allows access to higher mass region.
\end{itemize}

\acknowledgments

T.~L. is supported by the National Natural Science Foundation of China (Grant No. 12375096, 12035008, 11975129).

\bibliographystyle{JHEP}
\bibliography{refs}

\end{document}